\documentclass[letter]{aa} % for the letters 
\usepackage{relsize}
\usepackage{upgreek}
\usepackage{amsmath}

\usepackage{graphicx}
\usepackage{txfonts}
\def\1E0102{1E\,0102}
\def\oiii{[O\,\textsc{\smaller III}]}
\def\fexiv{[Fe\,\textsc{\smaller XIV}]}
\def\fexi{[Fe\,\textsc{\smaller XI}]}
\def\fex{[Fe\,\textsc{\smaller X}]}

\begin{document}

   \title{\fexiv\ and \fexi\ reveal the forward shock in SNR 1E\,0102.2-7219}
   \titlerunning{\fexiv\ and \fexi\ reveal the forward shock in 1E\,0102.2-7219}

   \author{Fr\'ed\'eric P.~A. Vogt\inst{1}\thanks{ESO Fellow} 
         \and
         Ivo R. Seitenzahl\inst{2,3,4}
         \and
         Michael A. Dopita\inst{2}
         \and
         Parviz Ghavamian\inst{5}
          }

   \institute{$^1$ European Southern Observatory, Av. Alonso de C\'ordova 3107, 763 0355 Vitacura, Santiago, Chile. \\  
              $^2$ Research School of Astronomy and Astrophysics, Australian National University, Canberra, Australia.\\
              $^3$ ARC Centre for All-sky Astrophysics (CAASTRO).\\
              $^4$ School of Physical, Environmental and Mathematical Sciences, University of New South Wales, Australian Defence Force Academy, Canberra, ACT 2600, Australia\\
              $^5$Department of Physics, Astronomy and Geosciences, Towson University, Towson, MD 21252, USA.\\
      \email{frederic.vogt@alumni.anu.edu.au}
}

   \date{Received March 9, 2017; accepted May 05, 2017}
 
  \abstract
   {} %leave it empty if necessary  
   {We study the forward shock in the oxygen-rich young supernova remnant (SNR) \1E0102.2-7219 (\1E0102 in short) via optical coronal emission from \fexiv\ and \fexi: emission lines which offer an alternative method to X-rays to do so.}
   {We have used the Multi-Unit Spectroscopic Explorer (MUSE) optical integral field spectrograph at the Very Large Telescope (VLT) on Cerro Paranal to obtain deep observations of SNR 1E\,0102 in the Small Magellanic Cloud. Our observations cover the entire extent of the remnant with a seeing limited spatial resolution of 0.7\arcsec$\equiv$0.2\,pc at the distance of \1E0102. }
   {Our MUSE observations unambiguously reveal the presence of \fexiv\ and \fexi\ emission in \1E0102. The emission largely arises from a thin, partial ring of filaments surrounding the fast moving O-rich ejecta in the system. The brightest \fexiv\ and \fexi\, emission is found along the eastern  and north-western sides of \1E0102, where shocks are driven into denser ISM material, while fainter emission along the northern edge reveals the location of the forward shock in lower density gas, possibly the relic stellar wind cavity.  Modeling of the eastern shocks and the photoionization precursor surrounding \1E0102, we derive a pre-shock density $n_H = (7.4\pm 1.5)$\,cm$^{-3}$, and a shock velocity $330\,\mathrm{km}\,\mathrm{s}^{-1}< v_s < 350$\,km\,s$^{-1}$.}
   {}


   \maketitle

%-------------------------------------------------------------------
% Start of the real thing:
\section{Introduction}
The supernova remnant (SNR) \object{1E\,0102.2-7219} \citep[\1E0102 in short;][]{Dopita1981,Tuohy1983} is located in the \object{Small Magellanic Cloud} (SMC). With an age of (2054$\pm$584)\,yr \citep[][]{Finkelstein2006}, it belongs to the class of oxygen-rich (O-rich) young SNRs. The characteristic fast-moving O-rich ejecta in this system --the stellar chunks left-over from the SN explosion-- form an intricate set of filaments \citep[both spatially and in velocity space;][]{Finkelstein2006,Vogt2010}. These ejecta are detected at optical wavelengths primarily via the \oiii$\,\lambda\lambda$4959,5007\AA\ lines \citep[but see also][]{Blair2000}, following their encounter at a few 1000\,km\,s$^{-1}$ with the reverse shock \citep[][]{Sutherland1995}. 

The forward shock wave triggered by the SN explosion, on the other hand, creates a shell of hot X-ray emitting gas surrounding the SNR, as it ploughs through the surrounding medium. This hot gas phase in SNRs is typically studied via X-ray observations, and the case of \1E0102 is no exception \citep[see e.g.][]{Hughes2000,Gaetz2000,Rasmussen2001,Flanagan2004}. But this hot phase in SNRs can also be detected via coronal emission such as \fexiv$\,\lambda$5303\AA, \fexi$\,\lambda$7892\AA\ and \fex$\,\lambda$6375\AA\ \citep[][]{Shklovskii1967}. Accessible at visible wavelengths, these forbidden lines can enable ground-based observations of the hot phase in SNRs at high spatial resolution. Perhaps more importantly, these coronal lines are also highly temperature sensitive \citep{Kurtz1972}, and thus can allow differentiation of the different structural layers of the forward-shocked medium, otherwise blended at X-ray wavelengths. Forbidden coronal Fe emission has already been detected in several SNRs \citep[][]{Woodgate1974,Murdin1978,Lucke1979,Woodgate1979,Dopita1979,Itoh1979}. Most recently, \cite{Dopita2016} used integral field spectroscopy to tie the \fexiv\ emission in SNR N49 to its blast wave, deriving associated shock velocities $v_s= (350-400)$\,km\,s$^{-1}$. In this letter, we report the first unambiguous detection and spatial mapping of both \fexiv\ and \fexi\ emission from the forward shock and associated denser cloud shocks in SNR \1E0102. 

%--------------------------------------------------------------------
\section{Observations, data reduction \& post-processing}\label{sec:obs}

SNR \1E0102 was observed with the Multi-Unit Spectroscopic Explorer (MUSE) in Service Mode at the Very Large Telescope on the night of October 7, 2016, as part of Director Discretionary Time (DDT) program 297.D-5058 (P.I.: F.P.A. Vogt). The observations are comprised of $9\times900$\,s exposures on-source, interleaved with 180\,s exposures targeting an empty sky field away from the SMC, located at $02^{h}07^{m}41^{s}.0$; $-72^{\circ}14^{\prime}43^{\prime\prime}.0$ [J2000].
The large angular separation ($\sim$4.9$^{\circ}$) between the object field and the sky field led us to create individual observing blocks (OBs) for each exposure, assembled in three OSOSO concatenations. These concatenations were acquired back-to-back over a $\sim$4~hr period, with 1E0102 below 1.6 airmass in all cases and with a stable seeing of 0.7$^{\prime\prime}$ measured in the SGS throughout. Small spatial shifts $\sim$0.7$^{\prime\prime}$ and position angle changes of 90$^{\circ}$ were applied to each object exposure to help remove the residual background artefacts associated with the 24 individual integral field spectrographs within MUSE.

Each on-target exposure was reduced individually and later combined into a single cube using the \textsc{reflex} \citep{Freudling2013} MUSE workflow (v1.6). The combined datacube corresponds to 8100\,s on-source, with a spatial full-width at half-maximum (FWHM) measured on stars in the field-of-view of $0.7^{\prime\prime}$ in the V-band. Spectrally, the cube extends from 4750\,\AA\ to 9350\,\AA\ in steps of 1.25\,\AA. The combined datacube has a size of 323$\times$326 spatial pixels (spaxels) $\equiv$ 64.6\arcsec$\times$65.2\arcsec $\equiv$ 19.4\,pc$\times$19.6\,pc, assuming a distance to the SMC of 62\,kpc \citep[][]{Graczyk2014,Scowcroft2016}. For consistency with the Second Digitized Sky Survey (DSS-2) R-band image of the area, the WCS coordinates of the datacube were shifted by $+6$ spaxels in the $x$-direction and $-0.5$ spaxels in the $y$-direction. Every spectrum in the datacube was corrected for Galactic extinction along the line of sight using the \cite{Schlafly2011} recalibration of the \cite{Schlegel1998} infrared-based dust map, which assumes a \cite{Fitzpatrick1999} reddening law with Rv = 3.1 and a different source spectrum than \cite{Schlegel1998}. This correction was performed using the \textsc{brutus} code \citep[][]{Vogt2017} with A$_\mathrm{B}$=0.134 and A$_\mathrm{V}$=0.101 extracted from the NASA Extragalactic Database.

%--------------------------------------------------------------------

The numerous stars across the MUSE field-of-view coupled to a spatially-varying nebular continuum complicate the extraction and analysis of the emission lines associated with \1E0102. Similarly to \cite{Vogt2016a}, we rely on the Locally Weighted Scatterplot Smoothing algorithm \citep[LOWESS;][]{Cleveland1979} to fit individually the continuum for each spatial pixel (spaxel) in the datacube. The LOWESS algorithm provides non-parametric fits that are robust against the presence of hot pixels and/or emission lines via an iterative fitting \& outlier rejection approach. However, the spectral width of emission lines associated with the fast-moving O-rich SNR ejecta along certain lines of sight imply that performing a single LOWESS fit is satisfactory for the majority of spaxels -- but not all. To ensure a reliable continuum fit for \textit{all} spaxels, we adopt the following procedure:
\begin{enumerate}
\item Perform an initial low-resolution \& robust LOWESS fit (using a spectral windowing size of 920\,\AA).
\item For each spectrum, crop all spectral pixels brighter than 5$\times$ the intensity level of the initial LOWESS fit. This effectively crops strong emission lines like \oiii$\,\lambda\lambda$5007,4959\AA\, which are much brighter than the continuum and affect LOWESS fits with smaller spectral windowing.
\item Perform a second LOWESS fit on the masked spectrum using smaller (sharper)  windowing size of 250\,\AA). 
\item Interpolate the second LOWESS fit over the masked pixels using Akima splines \citep{Akima1970}.
\end{enumerate} 

\begin{figure}[hb!]
\centerline{\includegraphics[scale=0.45]{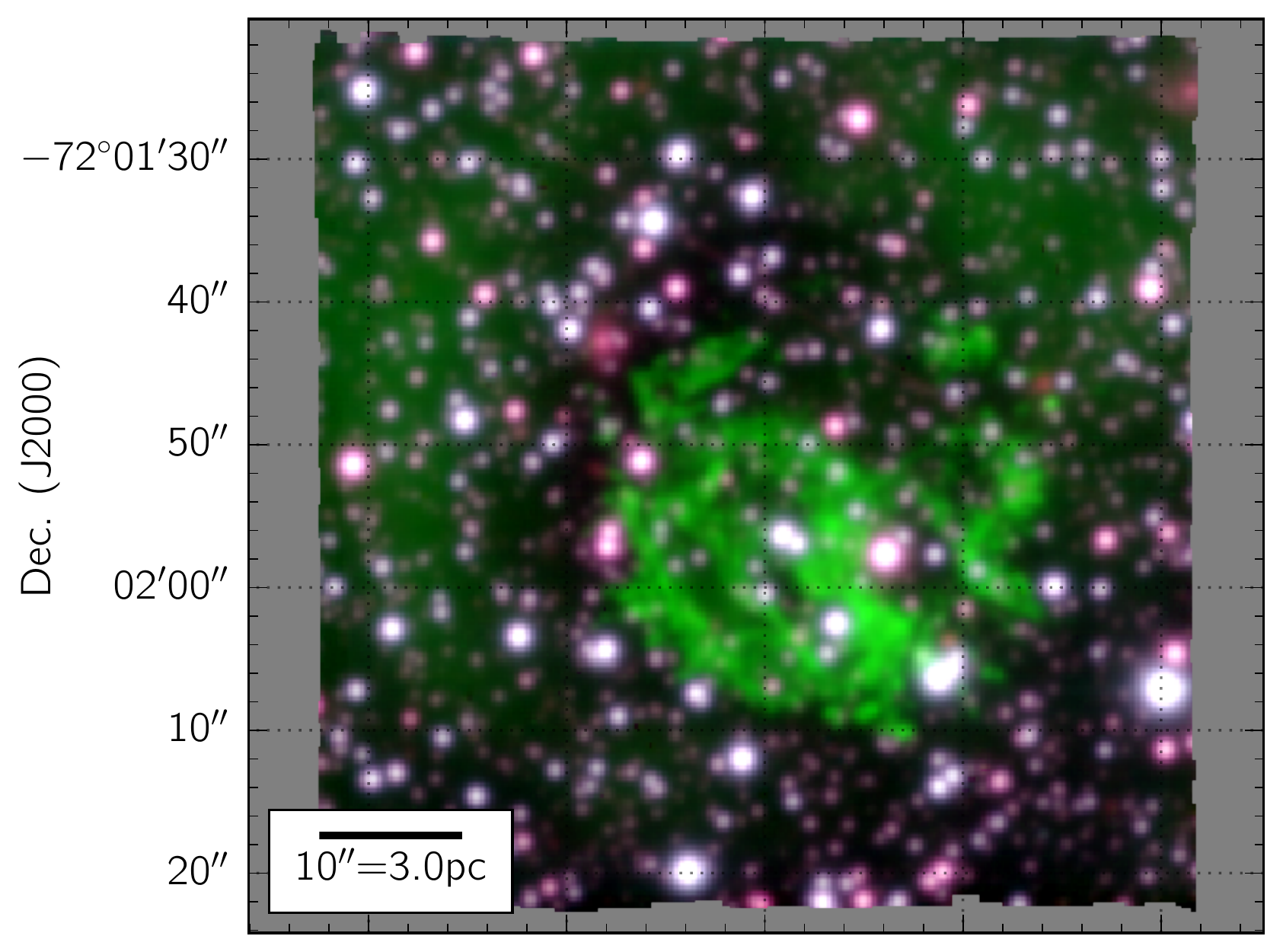}}
\centerline{\includegraphics[scale=0.45]{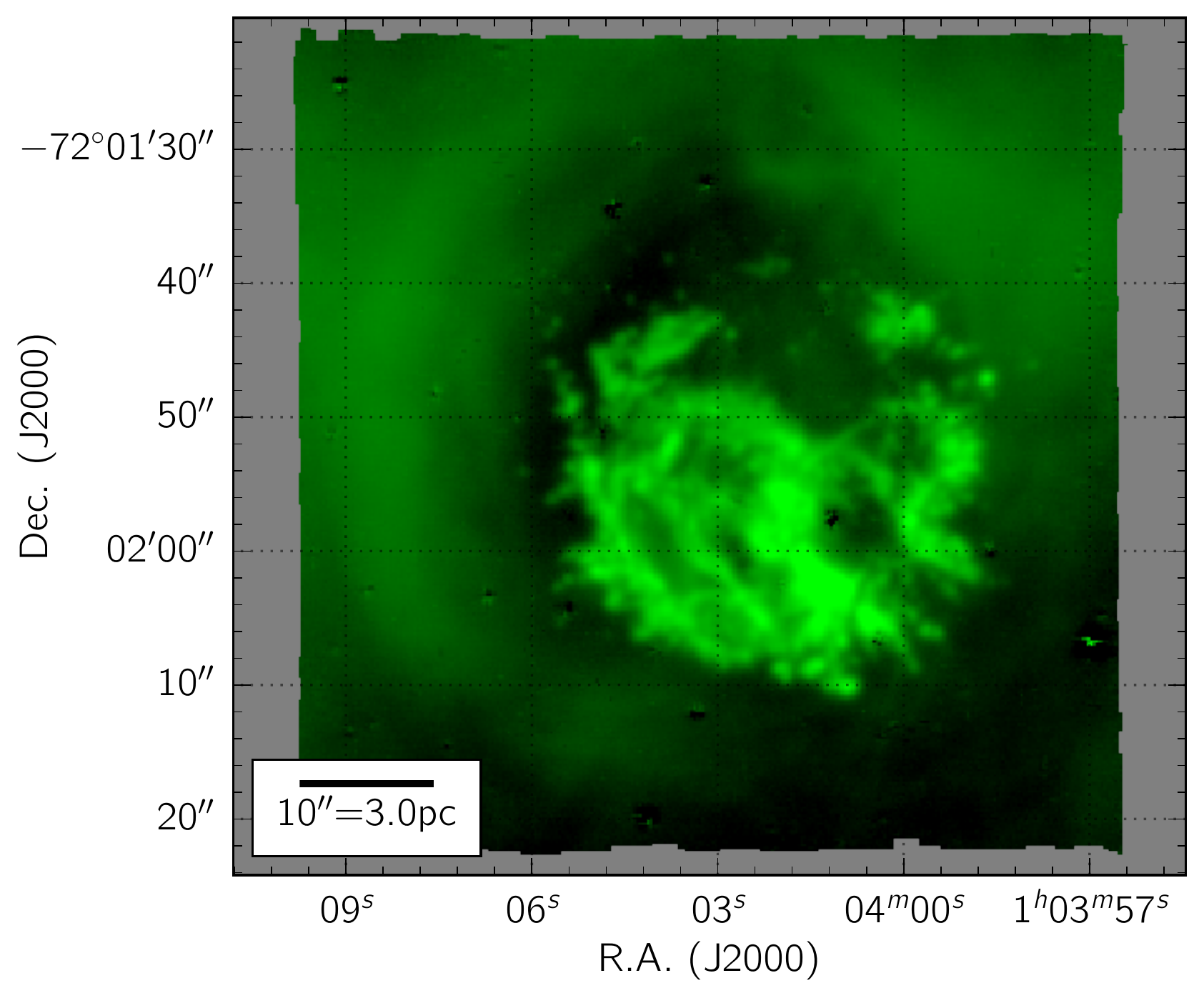}}
\caption{Pseudo-RGB images constructed from the combined MUSE datacube before (top) and after (bottom) removing the stellar and nebular continuum in each spaxel, with the R, G and B channels corresponding to the summed 6900\,\AA$\rightarrow$7100\,\AA, 4900\,\AA$\rightarrow$5100\,\AA, and 5300\,\AA$\rightarrow$5500\,\AA\ (observed) spectral ranges, respectively. The intensity stretch of each channel differs between both images for greater clarity. Stars appear in hues of blue and red, whereas the O-rich fast-moving ejecta are visible as bright green filaments.} \label{fig:MUSE_cont_sub}
\end{figure}

The efficiency of this procedure is illustrated in Fig.~\ref{fig:MUSE_cont_sub}, where we compare the MUSE datacube before and after continuum subtraction. The procedure is not perfect, and leads to residual artifacts associated with the brightest stars. For those, our continuum-removal procedure fails to account for the (sometimes numerous) absorption features in their high S/N spectra. For our present analysis, these artifacts do not significantly affect the spectral regions around \fexiv$\lambda$5303\AA, \fexi$\lambda$7891\AA\ and \oiii$\,\lambda\lambda$\,5007,4959\AA\ and will thus not be discussed further. 

\section{Results}

\begin{figure*}[htb!]
\centerline{\includegraphics[scale=0.4]{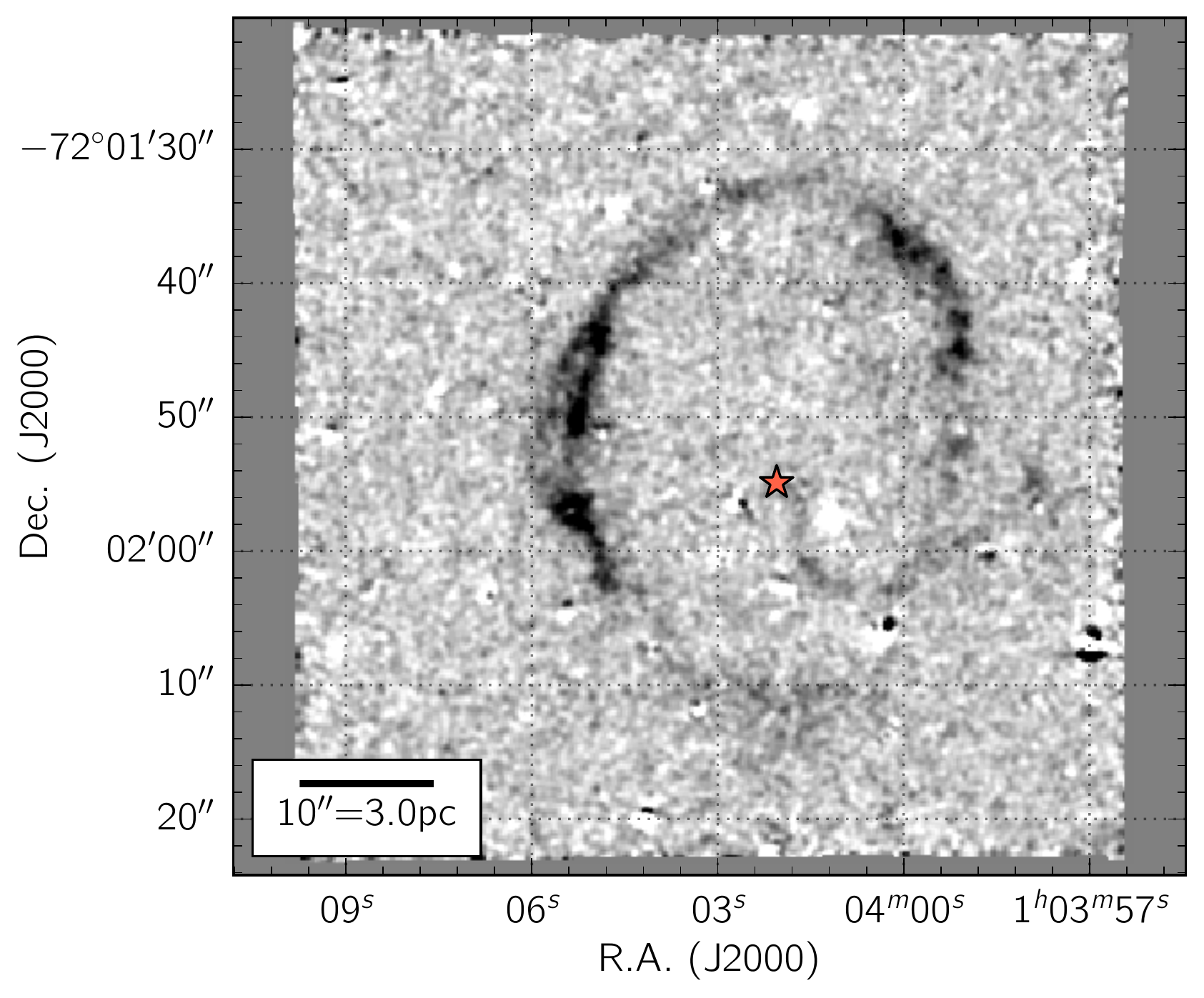}
            \includegraphics[scale=0.4]{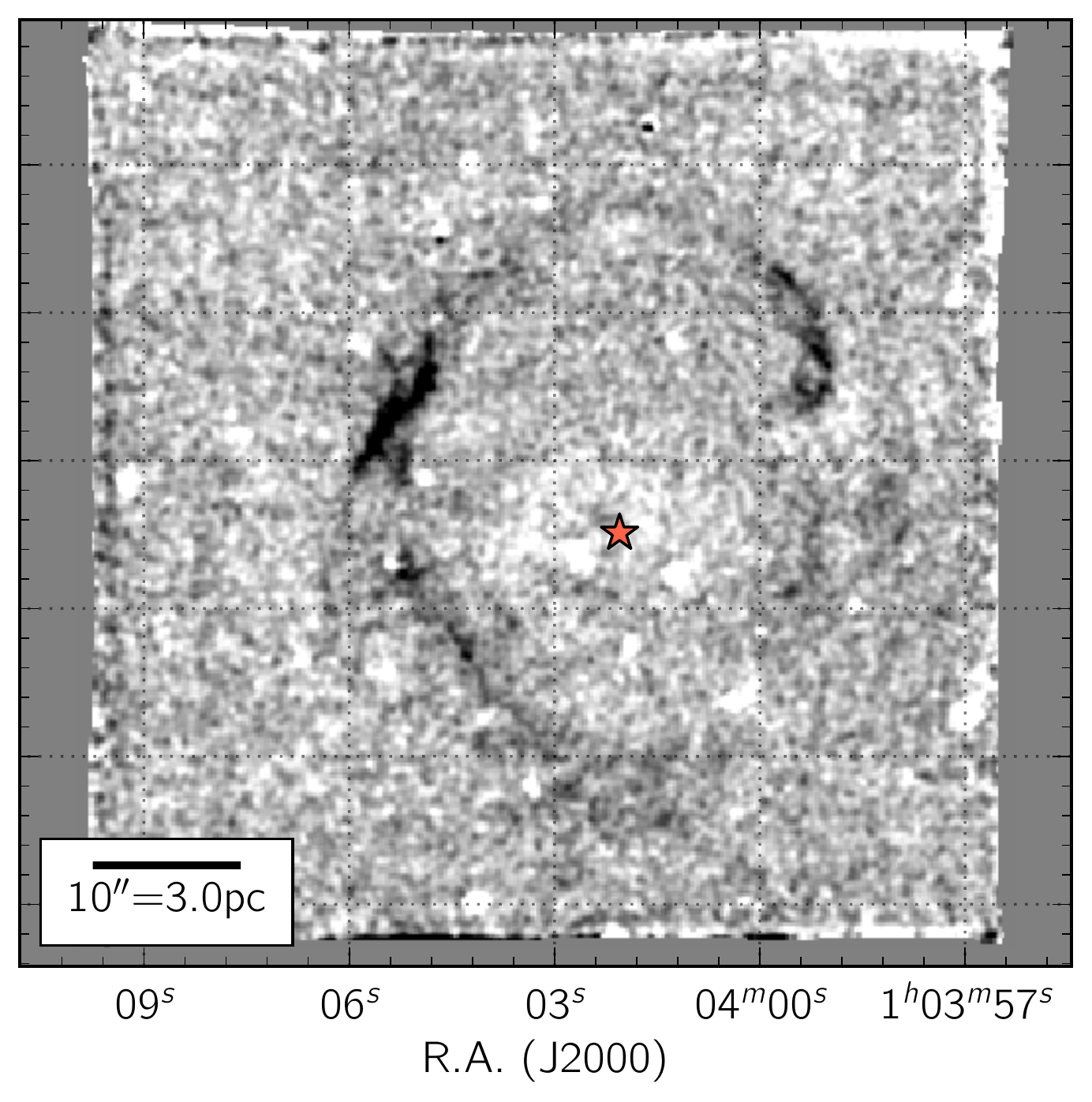}
            \includegraphics[scale=0.4]{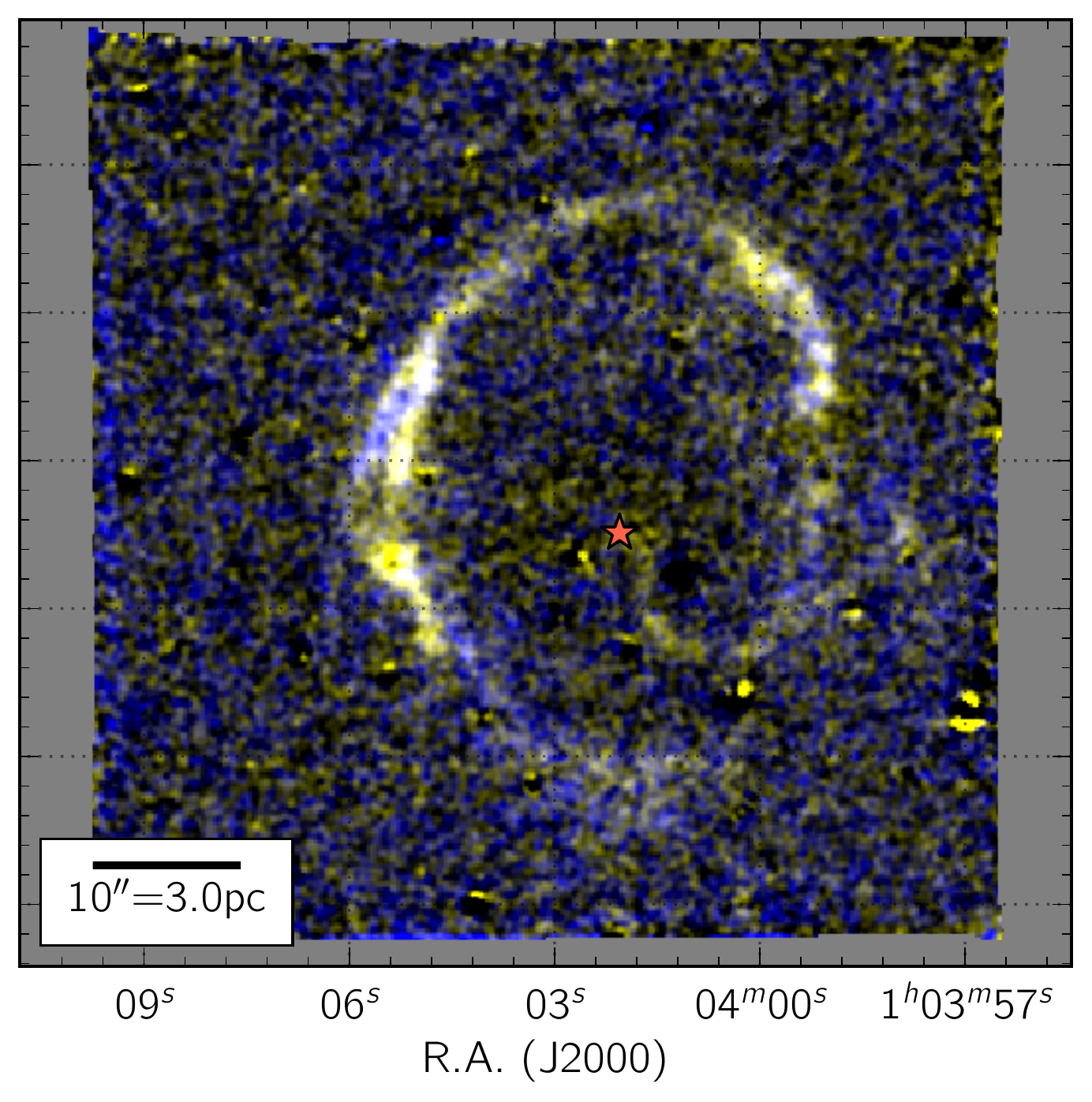}}
\caption{Left: integrated \fexiv\ emission. The image was smoothed using a Gaussian kernel with a smoothing scale of 1 spaxel across to enhance the fainter structures. The red star marks the proper motion center of the O-rich ejecta derived by \cite{Finkelstein2006}. Middle: idem, but for \fexi. Right: combination of both \fexiv\ (yellow) and \fexi\ (blue).  North is at the top and East to the left.}\label{fig:Fe-maps}
\end{figure*}

\begin{figure}[htb!]
\centerline{\includegraphics[scale=0.45]{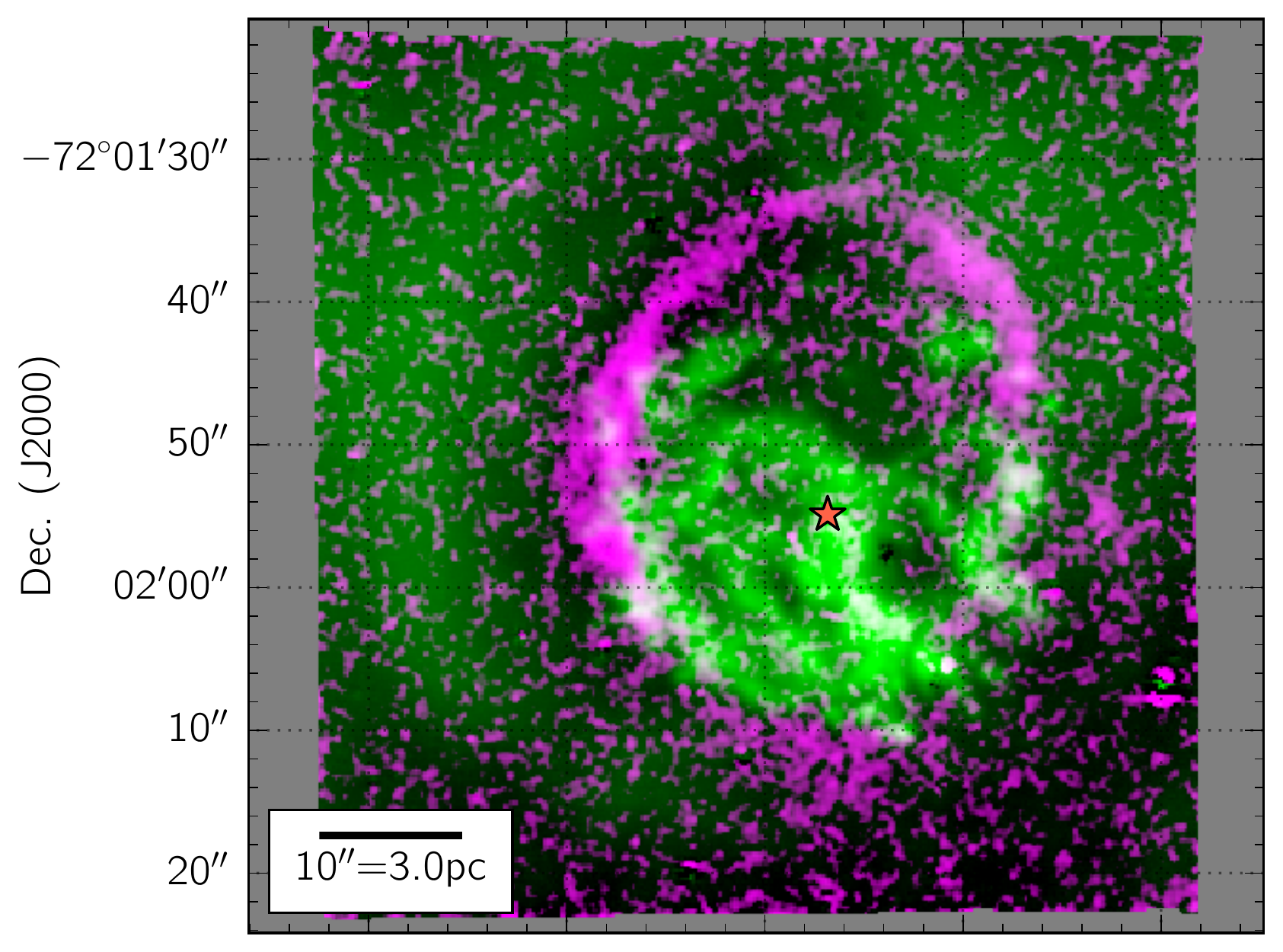}}
\centerline{\includegraphics[scale=0.45]{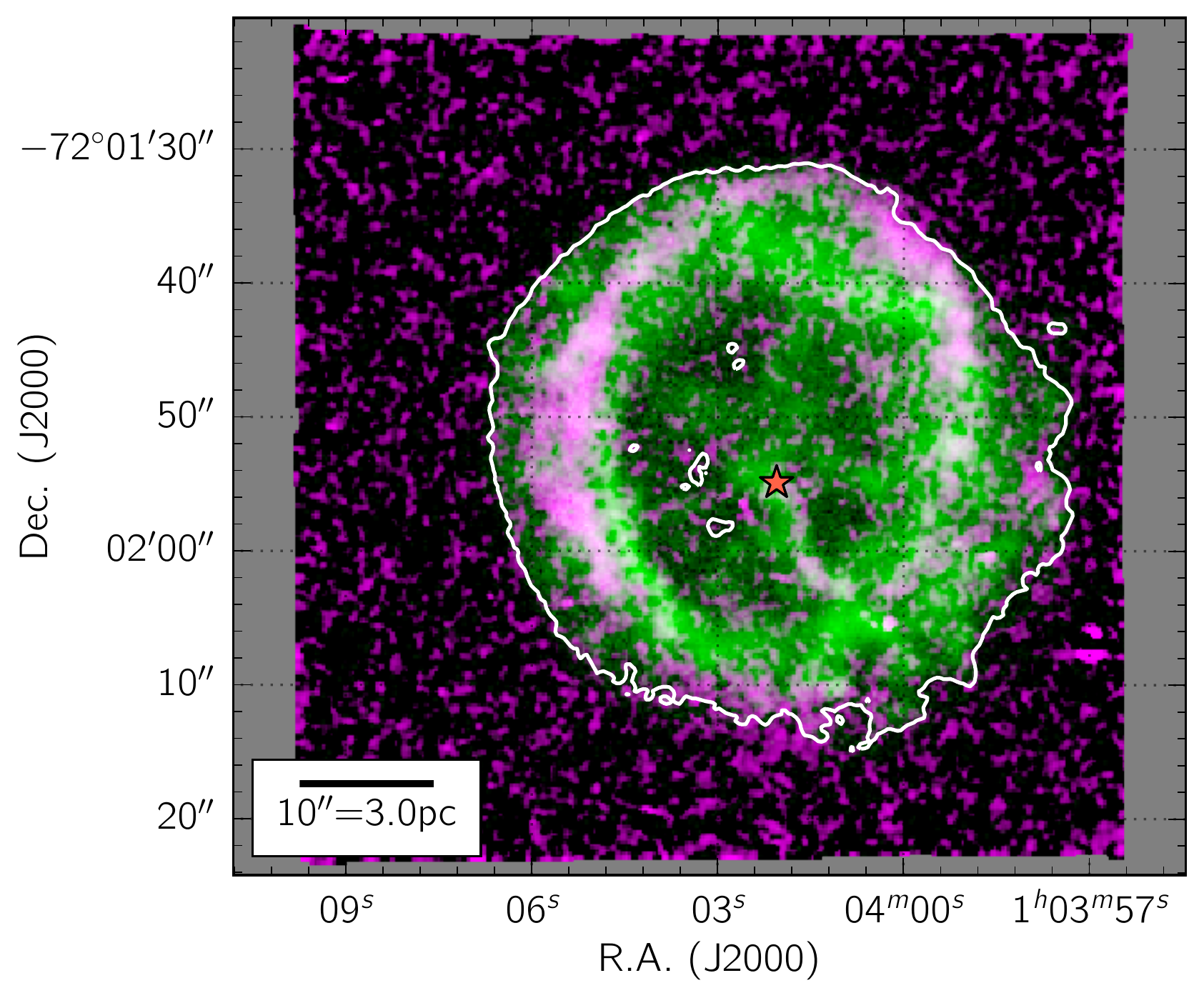}}
\caption{Top: comparison between the integrated \fexiv\ emission (purple) and integrated \oiii\ $\lambda\lambda$4959,5007\AA\ emission (green). The red star marks the proper motion center of the O-rich ejecta derived by \cite{Finkelstein2006}. Bottom: idem, but with the Chandra ACIS 1.1$\rightarrow$2\,keV X-ray flux instead [ACIS image credits: NASA/CXC/MIT/D.~Dewey et al. \& NASA/CXC/SAO/J.~DePasquale]. The full extent of the X-ray shell is traced with a white contour.}\label{fig:fe-oiii-xray}
\end{figure}

We present in Fig.~\ref{fig:Fe-maps} the integrated flux maps for the \fexiv\ and \fexi\ coronal lines, extracted from our continuum subtracted MUSE datacube. A thin ellipsoidal structure can be seen in these emission line maps, with brighter rims towards the East and the North-West. A faint arc predominantly visible in \fexiv\ is located West of the SNR center. Towards the South, the coronal emission is fainter but also appears more spatially extended. The \fexiv\ and \fexi\ emission encloses most of the O-rich ejecta and is contained (overall) within the X-ray shell of the SNR observed in Chandra ACIS observations\footnote{The Chandra ACIS data was downloaded from \url{http://chandra.harvard.edu/photo/openFITS/xray_data.html}.} (see Fig.~\ref{fig:fe-oiii-xray}). We marginally detect emission from \fex\,$\lambda$6375\AA\ co-spatial with \fexiv\ and \fexi, but the proximity of this line (in wavelength) to bright [O\,\textsc{\smaller I}]\,$\lambda$6300\AA\ emission from (redshifted) O-rich ejecta significantly complicates its analysis. We delay the discussion of \fex\ emission until a careful identification of all the fast moving ejecta in the datacube (outside the scope of this Letter) has been performed.
  
\subsection{Surface brightness of the coronal lines}\label{sec:S}
In the bright \fexiv\ and \fexi\ filament on the East side of the SNR, we measure an observed surface brightness $S_{\mathrm{[Fe\,\textsc{xvi}}],obs}= (1.4\pm0.2) \times 10^{-17}$\,erg\,cm$^{-2}$\,arcsec$^{-2}$\,s$^{-1}$ and $S_{\mathrm{[Fe\,\textsc{xi}}],obs} = (9\pm4) \times 10^{-18}$\,erg cm$^{-2}$\,arcsec$^{-2}$\,s$^{-1}$. At this location, the projected outer radius of the coronal line-emitting region is about 14.3\,arcsec or 4.3\,pc. The projected width of the filament is $\sim$0.5\,pc, which for a spherical shell implies that the flux adds over a path length through the filament of 4.4\,pc. The observed surface brightness of the shell is thus enhanced via limb-brightening by a factor $\eta\cong9$. Using an additional reddening correction of E(B-V)=0.04 to account for the SMC \citep[following][]{Blair2000}, implying a line flux attenuation A$_{\mathrm{[Fe\,\textsc{xvi}}]}=1.19$ and A$_{\mathrm{[Fe\,\textsc{xi}}]}=1.12$ according to the turbulent dust screen model from \cite{Fischera2005}, we thus derive the intrinsic surface brightness of the filament $S_\lambda=S_{\lambda,obs}\eta^{-1}A_\lambda$ to be: $S_\mathrm{[Fe\,\textsc{xvi}]}= (1.9\pm0.3) \times 10^{-18}$\,erg\,cm$^{-2}$\,arcsec$^{-2}$\,s$^{-1}$
and $S_\mathrm{[Fe\,\textsc{xi}]}= (1.1\pm0.5) \times 10^{-18}$\,erg\,cm$^{-2}$\,arcsec$^{-2}$\,s$^{-1}$.

\subsection{Inferred pre-shock density from the precursor emission}
A photoionisation precursor is very evident as a shell of diffuse \oiii\ emission in Fig.~\ref{fig:MUSE_cont_sub}. This shell, nearly 2 pc thick, is produced by the strong radiation field of the shocked O-rich ejecta in \1E0102.  From our MUSE data cube and previous observations with the WiFeS integral field spectrograph \citep{Vogt2016a}, we measure a mean H$\beta$ surface brightness in the northern precursor region (away from the obvious H{\,\textsc{\smaller II}} regions) of $S_{\mathrm H\beta }= (3.3\pm1.0) \times 10^{-16}$\,erg cm$^{-2}$ arcsec$^{-2}$ s$^{-1}$. The inner boundary of the precursor  is tightly defined by the size of the \fexiv\ shell (see Fig.~\ref{fig:fe-oiii-xray}): ${\sim}35$\,arcsec, or a radius of 5.1\,pc. The precursor's outer boundary is, on the other hand, not so well defined. If taken from the edge of the brightening in the \oiii\ shell, we  estimate it to have a diameter of ${\sim}52$\,arcsec, corresponding to a radius of ${\sim}7.8$\,pc. These figures imply a reddening-corrected total H$\beta$ flux for the precursor of  $F_{\mathrm H\beta }= (6.4\pm2) \times 10^{-13}$\,erg cm$^{-2}$ s$^{-1}$, which, correcting for the distance to the SMC, yields an absolute H$\beta$ luminosity for the whole precursor (treated as a complete shell) of $L_{\mathrm H \beta }= (2.4\pm0.8) \times 10^{35}$\,ergs s$^{-1}$. 

We can now estimate the mean density in the \oiii\ emitting region around \1E0102.  The luminosity in H$\beta$ is $ L_{\mathrm H \beta }= \epsilon _{\mathrm H \beta }Vn_H^2$, where $\epsilon _{\mathrm H \beta } = 1.014\times10^{-25}$erg cm$^3$ s$^{-1}$  for  Case B hydrogen emissivity \citep[][Table B.5, p.381]{Dopita2003}, $V$ is the ionised volume, and $n_H$ is the hydrogen density (cm$^{-3}$).  The resulting mean density in the precursor region from this calculation is $n_H = (7.4\pm 1.5)$\,cm$^{-3}$.

\subsection{Modelling the Blast Wave}
We have used the {\sc mappings v} code v10.3 to model the Fe forbidden emission line strengths and ratios for the bright filament on the eastern side of \1E0102. We used the SMC abundances from \cite{Russell1992}, with the heavy elements depleted onto dust according to the \cite{Jenkins2009} scheme. The base depletion factor was $\log D_{Fe} = -1.00$, which is appropriate for the diffuse ionised ISM. Our shock models have a reference pre-shock hydrogen density $n_H = 10$\,cm$^{-3}$ and are run for 2000\,yr: long enough for the ionization time-scale of the relevant Fe species to be short in comparison, so that we settle into a steady-state solution.

\begin{figure}[htb!]
\centerline{\includegraphics[scale=0.5]{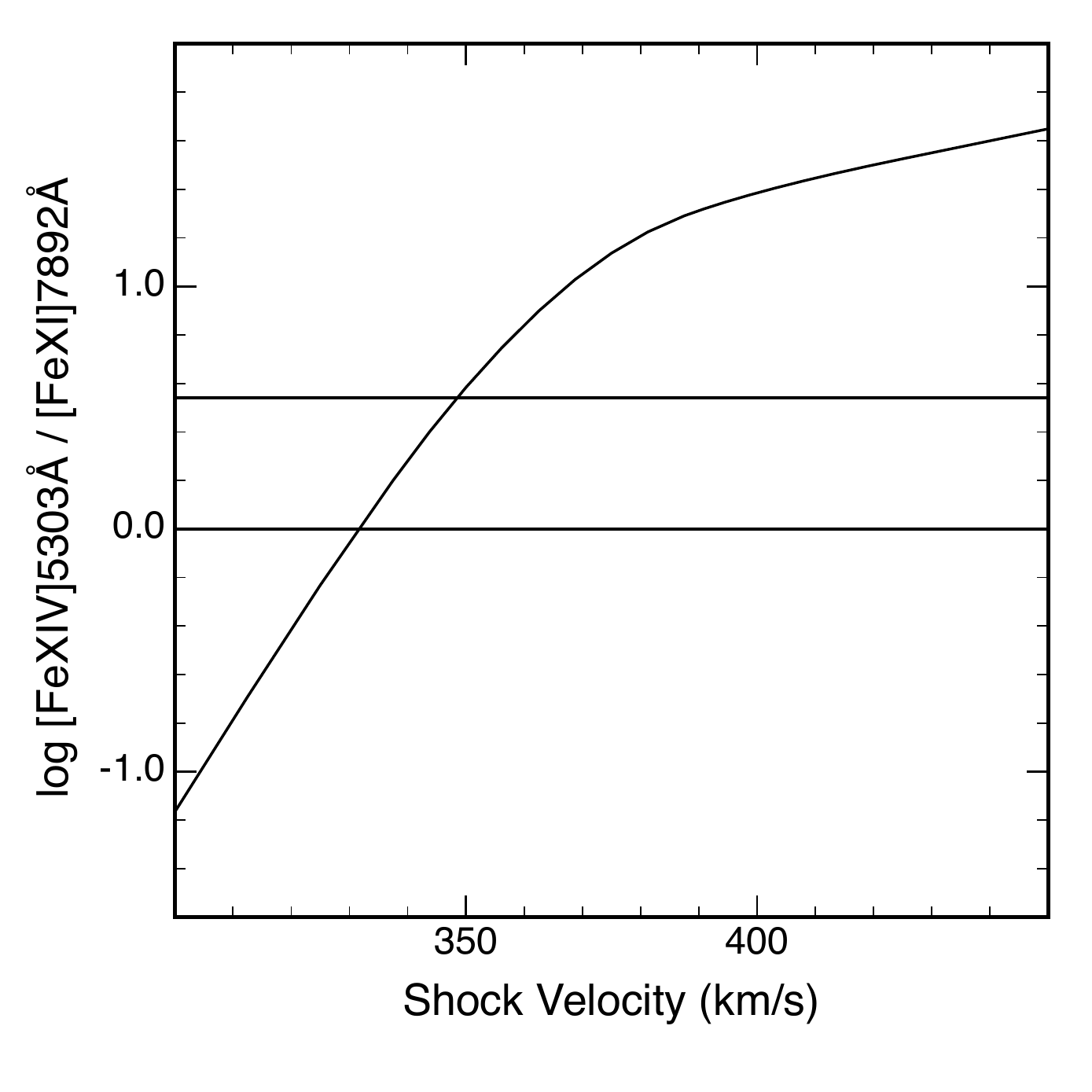}}
\caption{The computed [Fe\,\textsc{xiv}]5303\AA\ / [Fe\,\textsc{xi}]7890\AA\ ratio as a function of the blast wave velocity. The photometry of the filaments provides observed limits between the two horizontal lines, implying a blast-wave velocity of (330 -- 350) km s$^{-1}$.}\label{fig:Fe-ratio}
 \end{figure}

We find that the predicted flux ratio \fexiv\,$\lambda$5303\AA\ over \fexi\,$\lambda$7892\AA\ is very sensitive to the shock velocity and so provides an accurate way of determining this parameter (see Fig.~\ref{fig:Fe-ratio}). The observations put limits on this flux ratio of between 1.0 and 3.7, which therefore implies that the shock velocity of the eastern filament is in the range $330\,\mathrm{km}\,\mathrm{s}^{-1} < v_s  < 350$\,km\,s$^{-1}$. For the model with $v_s = 340$\,km\,s$^{-1}$, we infer that 
$S_{\mathrm{[Fe\,\textsc{\smaller xiv}]}}=  3.9 \times 10^{-18} (n_{10})(D_{Fe}/0.1)$\,erg\,cm$^{-2}$\,arcsec$^{-2}$\,s$^{-1}$, where $n_{10}$ is the assumed pre-shock hydrogen density in units of 10\,cm$^{-3}$ and $D_{Fe}$ is the assumed depletion factor of Fe in the ISM. Likewise, 
$S_{\mathrm{[Fe\,\textsc{\smaller xi}]}}=  3.3\times 10^{-18}(n_{10})(D_{Fe}/0.1)$\,erg\,cm$^{-2}$\,arcsec$^{-2}$\,s$^{-1}$. Comparing these numbers with those in Sec.~\ref{sec:S}, we infer that the pre-shock density for the eastern filament is $n_H=3-5$\,cm$^{-3}$: in reasonable agreement with the density estimated from the precursor luminosity above of  $n_H = (7.4\pm 1.5)$\,cm$^{-3}$.

The picture emerging from our analysis is that of a supernova remnant expanding into an asymmetric bubble swept out by the progenitor WR star winds. The bubble is narrower along the East-West direction (as noted by Finkelstein et al. 2006), and the blast wave of \1E0102 has struck denser gas filaments on these sides of the remnant. In doing so, the main shock wave moving outwards at a few 1000\,km\,s$^{-1}$ in a low density medium with n$_H$$\sim$0.1\,cm$^{-3}$, has generated the local, slower, non-radiative shocks with speeds of $\sim$340\,km\,s$^{-1}$ simulated above. Along the northern and southern edges, the SNR is still expanding through the wind cavity (traced by the dark \oiii\  hole seen in Figure 3).  In particular, the northern edge of the X-ray remnant fills most, but not all of the \oiii\ hole: a fact consistent with the propagation of the blast wave through a low density cavity.

\begin{acknowledgements}
We thank the anonymous referee for a prompt and constructive report. This research has made use of \textsc{brutus}, a Python module to process data cubes from integral field spectrographs hosted at \url{http://fpavogt.github.io/brutus/}. For this analysis, \textsc{brutus} relied on \textsc{statsmodel} \citep{Seabold2010},
\textsc{matplotlib} \citep{Hunter2007}, \textsc{astropy}, a community-developed core Python package for Astronomy
\citep{AstropyCollaboration2013}, \textsc{aplpy}, an open-source plotting package for Python hosted at \url{http://aplpy.github.com}, and \textsc{montage}, funded by the National Science Foundation under Grant Number ACI-1440620 and previously funded by the National Aeronautics and Space Administration's Earth Science Technology Office, Computation Technologies Project, under Cooperative Agreement Number NCC5-626 between NASA and the California Institute of Technology. 

This research has also made use of the \textsc{aladin} interactive sky atlas \citep{Bonnarel2000}, of \textsc{saoimage ds9} \citep{Joye2003} developed by Smithsonian Astrophysical Observatory, of NASA's Astrophysics Data System, and of the NASA/IPAC Extragalactic Database \citep[NED;][]{Helou1991} which is operated by the Jet Propulsion Laboratory, California Institute of Technology, under contract with the National Aeronautics and Space Administration. 

IRS was supported by Australian Research Council Laureate Grant FL0992131. PG thanks the Stromlo Distinguished Visitor Programme. FPAV and IRS thank the CAASTRO AI travel grant for generous support. We also thank Ashley Ruiter for helpful comments on the manuscript.

Based on observations made with ESO Telescopes at the La Silla Paranal Observatory under programme ID 297.D-5058[A].

\end{acknowledgements}

\bibliographystyle{aa} % style aa.bst
\bibliography{bibliography_fixed} % your references Yourfile.bib

\begin{appendix} %First appendix

\end{appendix}

\end{document}